# Angstrom Scale Ionic Memristors' Engineering with van der Waals Materials- A Route to Highly Tunable Memory States

Dhal Biswabhusan[1][†], Puzari Animesh[1][†], Li-Hsien Yeh[3,4], Kalon Gopinadhan*[1,2]

† = Equally contributed

[1]Department of Physics, Indian Institute of Technology Gandhinagar, Gujarat 382355, India

[2]Department of Materials Engineering, Indian Institute of Technology Gandhinagar, Gujarat 382355, India

[3]Department of Chemical Engineering, National Taiwan University of Science and Technology, Taipei 10607, Taiwan

[4]Advanced Manufacturing Research Center, National Taiwan University of Science and Technology, Taipei 10607, Taiwan

**\*Corresponding author Email** - gopinadhan.kalon@iitgn.ac.in

## Abstract

**Memristors that mimic brain functions are crucial for energy-efficient neuromorphic devices. Ion channels that emulate biological synapses are still in the early stages of development, especially the tunability of memory states. Here, we demonstrate that cations such as $K^+$, $Na^+$, $Ca^{2+}$, and $Al^{3+}$ intercalated in the interlayer spaces of vermiculite, result in highly confined channels of size 3-5 Å. They host exotic memristor properties through ion exchange dynamics, even at high salt concentrations of 1 M. The bipolar memristor characteristics observed are tunable with frequency, geometric asymmetry, ion concentration, and intercalants. Notably, we observe polarization-flipping memristor behavior in two cases: one with $Al^{3+}$ ions and another with devices having a geometric asymmetry ratio greater than 15. This inversion is attributed to the over-screening of counter-ions due to their accumulation at the channel entrance. Our results suggest that ion exchange dynamics, ion-ion interactions, and ion accumulation/depletion mechanisms, particularly with multivalent ions, can be harnessed to develop advanced memristor devices.**

## Introduction

The brain's structure and functionality, where memory and processing happen in a single place, is an excellent reference for designing energy-efficient neuromorphic devices[1,2]. Crucially, biological brains use ions for data processing, unlike electrons in a semiconductor processor[3, 4]. The solid-state version, the so-called memristors (memory with resistor) are analogous to biological neuron synapses[5, 6] that exhibit features such as (i) pinched current ($I$) – voltage ($V$) loop, forming two distinct conductance states that are accessible with a voltage, (ii) linear $I$-$V$ characteristics beyond a threshold frequency and (iii) frequency-dependent hysteretic *I-V* loop. Biological brains offer significant advantages over artificial electronic memristors as the former is flexible with several charge carriers ($K^+$, $Na^+$, $Ca^{2+}$, and $Al^{3+}$) and has a wide range of ionic conductivity, and each conductivity state can store information. Moreover, the conductivity of such biological synapses shows excellent tunability with various stimuli such as pH, voltage, humidity, temperature, pressure, and concentration gradient. Angstrom-sized ion channels in such systems are believed to be crucial in controlling charge transport. However, mimicking biological brains with solvated salt ions has still been a distant goal.

Nevertheless, there has been reasonable success in fabricating artificial fluidic channels in the last decade, especially channels of nanometer sizes or smaller. They demonstrated several interesting features, such as ultrafast water[7] and ion permeation[8] and selective transport of cation versus anion or vice versa[9, 10, 11]; however, simulating brain-like function remained challenging. Few groups demonstrated the ionic memristor effect with nanofluidic devices, especially low-dimensional systems such as nanopores[12, 13, 14, 15] and nanochannels[16, 17], pristine $MoS_2$, and activated carbon channels[18]. These devices generated great interest as they utilize mechanisms such as ion concentration polarization[16], ion adsorption/desorption[15, 18], and ionic liquid/salt solution interfacial movement[17] with memory time varying from milliseconds to hours.

However, in most cases, an exceedingly difficult fabrication process was used. Fabricating highly confined fluidic channels with angstrom-scale precision is still a distant goal. In the reported fluidic devices, the ionic memristor effect becomes insignificant at physiological concentrations of 100 mM or higher because of ion screening effects. Moreover, at high salt concentrations, pristine $MoS_2$ exhibits a unipolar memristor effect arising from the Wein effect, and the conductance state is solely dependent on the voltage amplitude and not its sign[18]. However, the brain's memory relies on both the signs, positive and negative, of the electrical signal, which is classified as bipolar memory, a key aspect when evaluating brain-like memory. Most solid-state devices show bipolar memory but have less energy efficiency and little tunability.

Realizing the ionic memristor effect at high salt concentrations is important for several reasons. For example, high concentration ensures high ionic conductivity with reduced memristor noise and being in large numbers, ions move faster, resulting in high-speed memristors. Maintaining stable memory states is expected to be easier in high concentrations, even if the environment is fluctuating. When the concentration is high, the probability of switching conductance states from ON to Off (or vice-versa) with a lower energy threshold is enhanced, resulting in energy-efficient devices.

Here, we design a spatially asymmetric angstrom size fluidic channel consisting of vermiculite laminates with a tunable interlayer spacing of $\sim 3 - 5$ Å, by intercalating hydrated ions of $K^+$, $Na^+$, $Ca^{2+}$, and $Al^{3+}$ in Li-vermiculite. These devices exhibit robust and tunable memristor behavior at intermediate and higher concentrations for several salt solutions. An apparent bipolar memory effect was observed with a low-frequency voltage signal. Our cost-effective fabrication technique allowed us to achieve a high asymmetric ratio (between the wide and the narrow entrance of the channel) from 8 to 25, resulting in clearly distinguishable ON/Off conductance states, even at 1 M concentration. Our chosen material has several advantages, like simple salt intercalation for membrane fabrication instead of harsh chemicals, earth abundance, cost-effectiveness, high surface charge, and high ion exchange capacity among all the clay materials[19].

**Results and Discussion**

Vermiculite is a class of 2:1 magnesium aluminosilicate clay, and one layer of vermiculite consists of 3 sheets; one sheet, occupied by $Al^{3+}$ ions at the octahedral sites, sandwiched between two sheets of $Si^{4+}$ ions at the tetrahedral sites. However, some $Si^{4+}$ tetrahedral sites are substituted by $Al^{3+}$ ions, leaving a net negative charge on the vermiculite surface. This charge is balanced by unintentional intercalation of cations such as $Mg^{2+}$ ions inside the interlayer spaces[20, 21]. We intentionally exchanged these interlayer cations with hydrated $K^+$, $Na^+$, $Ca^{2+}$, and $Al^{3+}$ ions, which tuned the accessible interlayer space to $\sim$3 Å, $\sim$5 Å, $\sim$5 Å, and $\sim$5 Å, respectively determined from XRD analyses (Figure S1). We denote the membranes of vermiculite intercalated with potassium as K-V, sodium as Na-V, calcium as Ca-V, and aluminium as Al-V.

**Easy fabrication technique to produce highly asymmetric channel with angstrom size fluidic memristor:** Our devices were fabricated, first by synthesizing a vermiculite dispersed solution, followed by preparation of 2D vermiculite membranes via vacuum filtration. These membranes were subsequently intercalated with alkali metal ions by immersing in 1 M aqueous solutions of KCl, NaCl, $CaCl_2$, and $AlCl_3$.

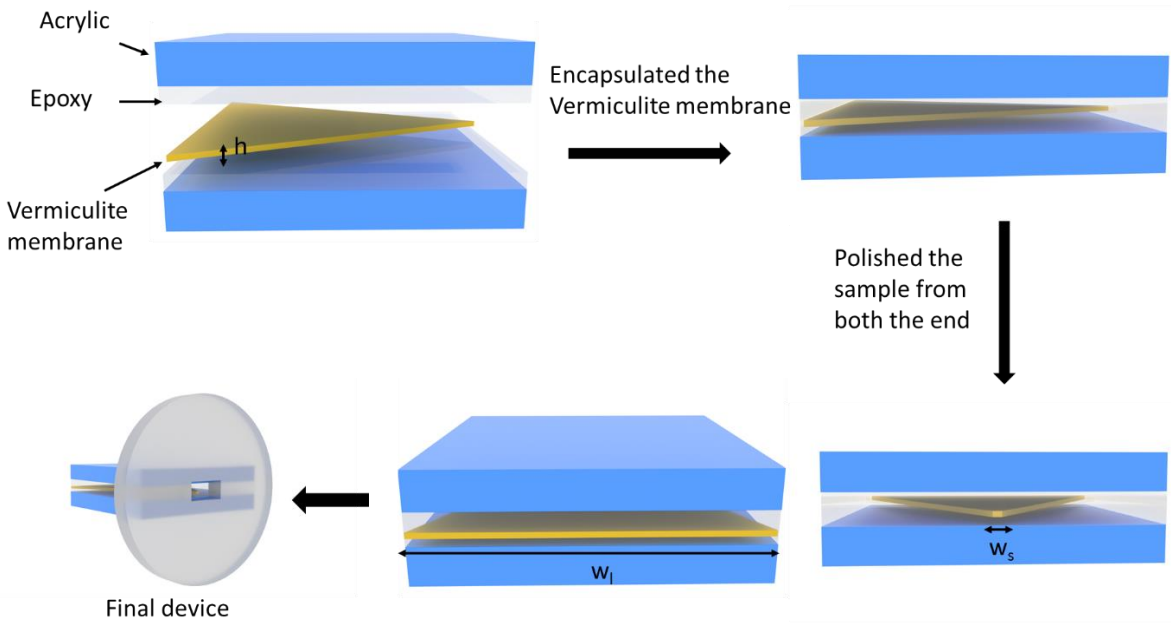

**Figure 1.** Device fabrication steps for ionic memristor transport studies across the vermiculite membrane with angstrom-sized capillaries. Here, $h$ is the height of the membrane, $w_l$ is the width of the larger entrance, and $w_s$ is the width of the smaller entrance. Images are not to scale.

Cuboidal-shaped acrylic structures (6 mm in length, 4 mm in width, and 3 mm in breadth) serve as the foundation for the memristor devices (Figure 1). The 2D vermiculite membrane was cut into isosceles triangles, then sandwiched between two acrylics with an epoxy (Loctite Stycast 1266). After 24 hours of drying under ambient conditions, both the entrances were gently polished with a sandpaper, and the final asymmetric entrance ratio was estimated using an optical microscope (Figure S2a). The triangular shape of the vermiculite membrane inherently provides an asymmetry between the entrances. One side of the exposed membrane had a larger area of $h \times w_l$ ($h = \sim$5 μm (Figure S2b), and $w_l$ varies from 4.5 to 5.5 mm), and another side, a smaller area of $h \times w_s$ ($w_s$ varies from 200 μm to 600 μm). In our devices, we restricted the asymmetry ratio ($AR = w_l/w_s$) in the range of 8 to 25; however, our fabrication technique is capable of going beyond this range. This straightforward and yet effective fabrication procedure provided a versatile platform to utilize geometric asymmetry in 2D memristor devices. The asymmetric sample was epoxy glued on an acrylic with a pre-drilled hole of 4 mm × 2 mm, for ion transport studies. This ensured that the path for the ion transport was only through our V-laminates (Figure 1). The length of the final membrane was 2.5 mm.

We placed this device between two reservoirs containing salt solutions of the same concentration. We chlorinated Ag electrodes to measure the ionic conductance. For the transport study with a salt ion, we used the corresponding salt-exchanged membranes, for example, KCl salt with K-V membranes, to avoid any ion exchange during the measurement.

Having fabricated high-quality asymmetric devices, we first studied the transport of KCl ions through K-V devices as these devices offer an extremely confined space of 3 Å and a large negative Zeta potential of -48 mV (Figure S3). We applied a voltage from -2 V to +2 V (direction of voltage; 0 → +2 V → 0 V → -2 V → 0 V) with a voltage step of 50 mV. Figure 2a shows the current ($I$)-voltage ($V$) characteristics at 100 mM concentration with a frequency of 0.4 mHz. Surprisingly, a pinched hysteresis loop was observed even at such high ionic concentrations. The inset of Figure 2a shows the conductance (=V/I) plotted against the applied voltage. Two distinct conductance states were seen, with the high (low) conductance state on the negative (positive) voltage side. However, its accessibility crucially depends on the history of the voltage applied. Interestingly, the hysteretic loop was very prominent only in mHz frequencies, and the loop area decreased with an increase in frequency (Figure 2a). At 50 mHz, a linear $I$-$V$ characteristic was seen without any hysteresis. The three fingerprints of a memristor as per Chua's theory[6], i.e., pinched hysteresis $I$-$V$ loop, reduction in the loop area with an increase in frequency, and the linear $I$-$V$ characteristics beyond a critical frequency were observed for K-V devices, suggesting that cation-intercalated vermiculite membranes are true ionic memristors.

The observed characteristics are contrastingly different from a capacitor holding stationary charges, in which the loop area increases with an increase in frequency. The memristor characteristics, therefore, clearly arising from dynamic charges. Vermiculite is a cation-exchange membrane and with K-V membranes in KCl solutions, there exists an intercalation and de-intercalation process. We believe this affects the ion transport dynamics and is responsible for the memristor characteristics. The charges are dynamically exchanged with interlayer cations[22, 23, 24], and in the case of K-V, the measured zeta potential is -48 mV, which is the average net charge of the surfaces in equilibrium with the solution. The negative zeta potential implies that cations are energetically favored for intercalation; however, the de-intercalation process is slow due to attractive interaction between the negatively charged surface and the cations. The asymmetric membrane geometry is crucial to see the memristor characteristics by accumulating and depleting charges at the entrance/exit, and the effect gets weaker with small asymmetry factors. We further verified our proposed mechanism by performing several experiments, which are discussed below.

The memristor behavior in vermiculite was further verified with sodium-exchanged vermiculite (Na-V) membranes and NaCl salt solutions. Figure 2b shows the $I$-$V$ studies performed at different frequencies for a sample of geometric asymmetry ratio (AR) 8. These samples also show true memristor characteristics, just like K-V samples. The role of confinement and surface charge on the memristor effect is further investigated with the help of Ca-V and Al-V samples. Figure 2c shows the memristor characteristics of the Ca-V membrane at 50 mM CaCl$_2$ solution. In this case also, the characteristics are similar to K-V samples. The van der Waal's gap in our Ca-V membrane is approximately 5 Å and the measured zeta potential is -16 mV (Figure S3). The influence of surface charge at a particular concentration, $C$ is better understood from the estimate of Debye length as given by eq 1;

$$\lambda_D = \sqrt{\frac{\varepsilon RT}{\sum_{i=1}^{N} F^2 Z_i^2 C_{i,0}}} \qquad (1)$$

Here, $\varepsilon$ is the permittivity of the solution, $R$ and $T$ are gas constants and temperature, respectively; $F$ is the Faraday constant, $N$ is the total number of ionic species, $Z_i$ and $C_{i,0}$ are the valence and bulk concentration of the $i^{th}$ ionic species, respectively. For Ca-V samples with $CaCl_2$ solution, at a chloride concentration of 100 mM, the Debye length is $\sim 6$ Å, which implies an overlapped Debye layer in our channels, leading to cation-selective transport. We tried to calculate the memory or diffusion time arising from the $Ca^{2+}$ transport by considering its diffusion coefficient and the geometric asymmetry. The diffusion time ($\tau$) for an asymmetric nanofluidic channel strongly depends on the length of the channel, $L$, the diffusion coefficient of the ions, $D$, and a factor due to geometric asymmetry, $\xi$ (1/AR), which is, $\tau = \frac{L^2}{4D}\xi$ [Ref.[2]]. For the Ca-V sample (Figure 2c), $L$ = 2.5 mm, $D$ = 0.8×10$^{-9}$ m²/s (Diffusion coefficient of $Ca^{2+}$ ion), and the value of $\xi$ is 0.083 which estimates a diffusion time of 162 sec and the corresponding frequency ($1/\tau$) is 6.1 mHz. The estimated value agrees well with our observations as the memristor behavior is mostly seen in the mHz range. By controlling the channel parameters, the memory time scale can be further tuned[2, 25].

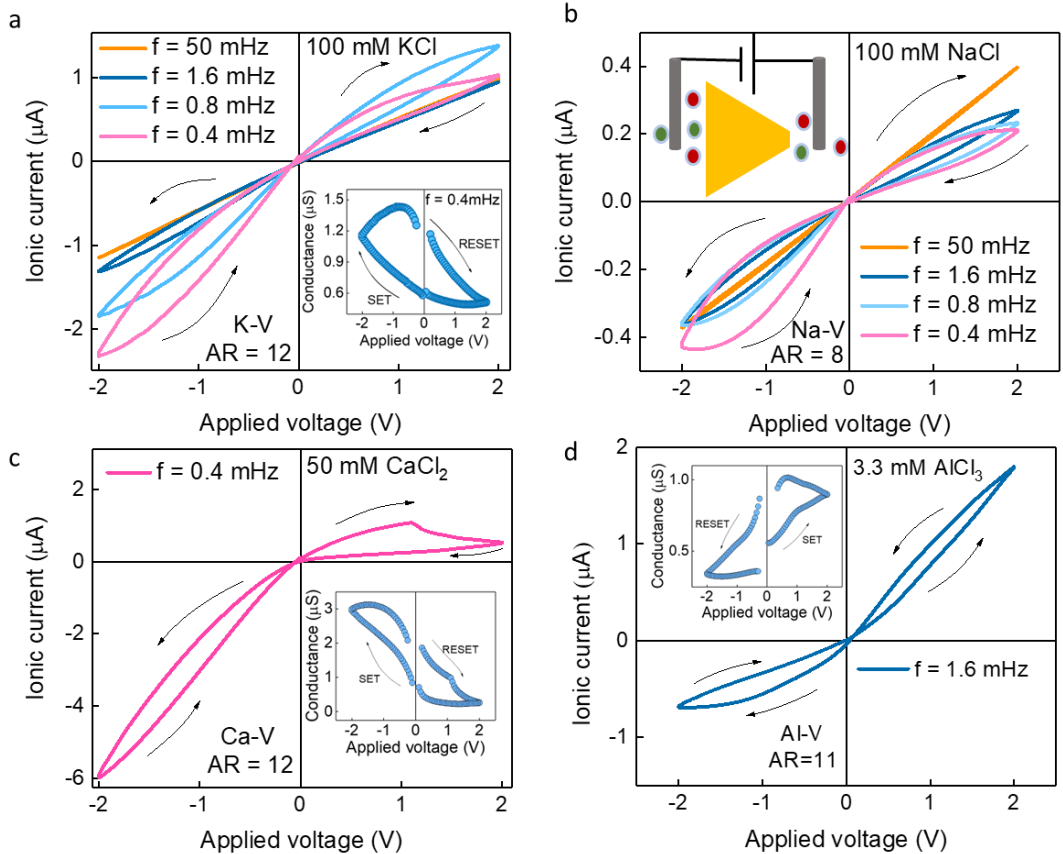

**Figure 2.** Memristor characteristics of asymmetric angstrom fluidic channels in intercalated vermiculite laminates. (a) The *I-V* curves at various scanning frequencies for a K-V device with a 100 mM KCl solution and an asymmetric ratio (AR) of 12. The arrow indicates the direction of the *I-V* loop. Inset: Conductance extracted from the *I-V* curve (Figure 2a, for f = 0.4 mHz) plotted as a function of applied voltage. (b) *I-V* curves

for a Na-V device at different scanning frequencies with a 100 mM NaCl solution and an AR of 8. Inset: Schematic of the ion transport measurement setup. (c) *I-V* curve for a Ca-V device measured at a frequency of 0.4 mHz with a 50 mM $CaCl_2$ solution and an AR of 12. The arrow indicates the direction of the *I-V* loop. Inset: Conductance extracted from the *I-V* curve plotted against the applied voltage. (d) *I-V* curve for an Al-V device at a frequency of 1.6 mHz with a 3.3 mM $AlCl_3$ solution and an AR of 11. The arrow indicates the direction of the *I-V* loop, which is flipped in the case of $AlCl_3$. Inset: Conductance extracted from the *I-V* curve plotted against the applied voltage.

In contrast, Al-V samples show memristor characteristics only at low concentrations of $AlCl_3$, i.e. 3.3 mM concentrations and lower. Although the van der Waals gap is similar to Ca-V membranes, the Al-V layers have little surface charge due to nearly complete ion exchange, resulting in poor ion selectivity, especially at higher concentrations. Surprisingly, in Al-V samples, the *I-V* curves are reversed, in which the high-conductance state flips to low-conductance, and vice versa (Figure 2d). The flipped (inverted) *I-V* loop and the appearance of memristor behavior only for low concentrations indicates that layers are getting a net positive charge.

Having established the role of confinement, we now examine the influence of geometric asymmetry in the memristor characteristics with vermiculite samples. We observed that the pinched *I-V* loop gets inverted when the asymmetric ratio of the device increases beyond 15 for K-V, Na-V and Ca-V samples. For samples with AR < 15 (Figure 2a), the high (ON) conductance state is reached when a sufficiently large negative voltage is applied; however, for samples with AR > 15 (Figure 3a & Figure S5a), the ON state is accessible only with a positive voltage. In contrast, Al-V samples showed an inverted *I-V* loop, irrespective of any AR (Figure 3c), and the ON conductance state is accessible only with a positive voltage. There are few instances of ionic current rectification (ICR) wherein a conductance reversal is reported as a result of a change in the type of surface charge due to solution pH[26] or when multivalent ions change the nature of surface charge via physical adsorption[27] . Very recently, an asymmetric channel with a pH gradient showed a reversal of ionic current rectification characteristics due to charge inversion[28]. Another study on a positively charged asymmetric nanopore indicated anions' transport over cations and showed higher conductance on the positive applied voltage side[15]. These studies at least hint that higher conductance can be achieved for positive applied potentials if anions are transported instead of cations. However, how is that a cation-selective channel transport anion? In our case, we found at least two experimental conditions where the *I-V* loop is inverted: (a) when we increased the AR beyond 15 in the case of K-V samples (also valid for Na-V and Ca-V samples) and (b) with Al-V samples.  Anions are suspected to be the reason for the observed behavior.

What mechanism can cause a dominant anion transport in high AR K-V samples? How exactly does the geometric asymmetry switch the conductance state in our membranes? The inset of Figure 3a shows the schematic of the measurement configuration. First, we consider devices with a small asymmetric ratio (AR) of less than 15 (For K-V, Na-V, and Ca-V). When a negative potential is applied to the narrow entrance with a wide entrance at zero potential (Reverse biasing), $K^+$ ions experience a negative potential and drift towards the narrow entrance from the wider entrance, resulting in enhanced conductance. $Cl^-$ experiences a repulsive force, and K-V being a cation selective channel, $Cl^-$ permeation would be very weak. Under forward biasing, $K^+$ ions depleted from the channel, resulting in the decrement of conductance. This is clear from the conductance vs. voltage plot provided in the inset of Figure 2a for low AR samples. To further increase the AR of K-V samples, we made the narrow entrance smaller. The *I-V* characteristics show a flip when compared to low AR samples. Due to asymmetric channels, naturally, $K^+$ ions accumulate at the narrow entrance. With the increase in concentration, there is a high probability of over-screening of negative surface charge at the narrow entrance, which changes the charge sign to positive and behaves as an anion-selective channel. A schematic depicting the mechanism is provided in Figure 3b. It is to be noted that the ionic conductance in asymmetric channels is completely determined by the narrow entrance. So, if the surface above the narrow entrance changes sign to positive, it selectively allows anions. Above a particular concentration, it is observed that the sign could be reversed (Figure S4a-d). Further, the flipped *I-V* is verified with NaCl salt with high AR (Figures S4a-d, S5b & S6a-b).

In contrast, we see inverted *I-V* characteristics in low AR Al-V samples (Figure 3c). This could be due to the charge inversion effect resulting from ion-ion interaction between multivalent ions at the surface. The charge on a bare surface could be inverted if the accumulation of counter-ion brings more charge than it requires to satisfy the electro-neutrality condition. Multivalent ions can form a two-dimensional layer close to the surface, and the charge density of that layer ($\sigma_{Stern}$) is referred to as the stern layer. For charge inversion to happen, the $\sigma_{Stern}$ must be greater than the bare surface charge density ($\sigma_b$) of the vermiculite. This is possible above a certain concentration called threshold concentration. We calculated the threshold concentration using the strongly correlated liquid (SCL) theory[27, 29, 30]. However, this is valid only for multivalent ions.

The threshold concentration is calculated using eq 2;

$$C_{threshold} = \left| \frac{\sigma_b}{2r_{ion}Ze} \right| exp\left( \frac{\mu_c}{k_B T} \right) \tag{2}$$

Where $r_{ion}$ is the hydrated ionic radius, $\sigma_b$ is the bare surface charge density, $Z$ is the ion valence, $e$ is the elementary charge of an electron, $k_B$ is the Boltzmann constant, $T$ (= 298 K) is the temperature, and $\mu_c$ is the chemical potential due to correlation effects which is given by eq 3.

$$\mu_c = -k_B T(1.65\Gamma - 2.61\Gamma^{0.25} + 0.26ln\Gamma + 1.95) \tag{3}$$

Here $\Gamma$ is the Coulomb coupling constant, the interaction parameter between the multivalent ions in the Stern layer (eq 4),

$$\Gamma = \frac{1}{4k_B T\varepsilon_0\varepsilon_r} \sqrt{\left| \frac{e^3 Z^3 \sigma_b}{\pi} \right|} \tag{4}$$

where $\varepsilon_r$ is the dielectric constant, the permittivity of free space. This theory is valid when $\Gamma \gg 1$, which is fulfilled for $Z > 3$, while the correlation effect is negligible for $\Gamma \ll 1$. From the Gouy-Chapman equation (Supplementary Note 1), we estimated the surface charge density, which is -4 mC/m$^2$ and assuming $\varepsilon_r = 80$, $\Gamma$ estimates to 0.2 for KCl ($2r_{ion}$=0.66 nm), and 1.02 for AlCl$_3$ ($2r_{ion}$=0.96 nm)[31]. For AlCl$_3$, $\Gamma$ is close to 1 but the theory is valid for $\Gamma \gg 1$. The value of $\Gamma$ depends on the surface charge of the sample. As in the case of vermiculite, K$^+$ or Li$^+$ ions are already adsorbed on the surface, and the zeta potential that we measured is -48 mV. However, in the case of GO and MXene systems, the zeta potential represents the bare surface potential due to physical adsorption. In the case of vermiculite, the measured zeta potential could be less compared to the bare surface potential due to the intercalation of the cations and also the chemical adsorption since the cations are part of the unit cell. In that case, the $\Gamma$ value would be increased for Z = 3 as the bare surface charge increases. So, it is possible to see the charge inversion effect with AlCl$_3$ in vermiculite, with a smaller asymmetric ratio and above a threshold concentration. From the calculation, we found that the threshold concentration for charge inversion is ~5 mM for AlCl$_3$, above which it can change the surface charge from negative to positive. The positive surface charge on vermiculite will allow selective transport of anions which explains the inverted *I-V* curve in the case of Al-V samples at 3.3 mM concentration, in agreement with the theoretical calculations. A schematic representation of the mechanism is provided in Figure 3d.

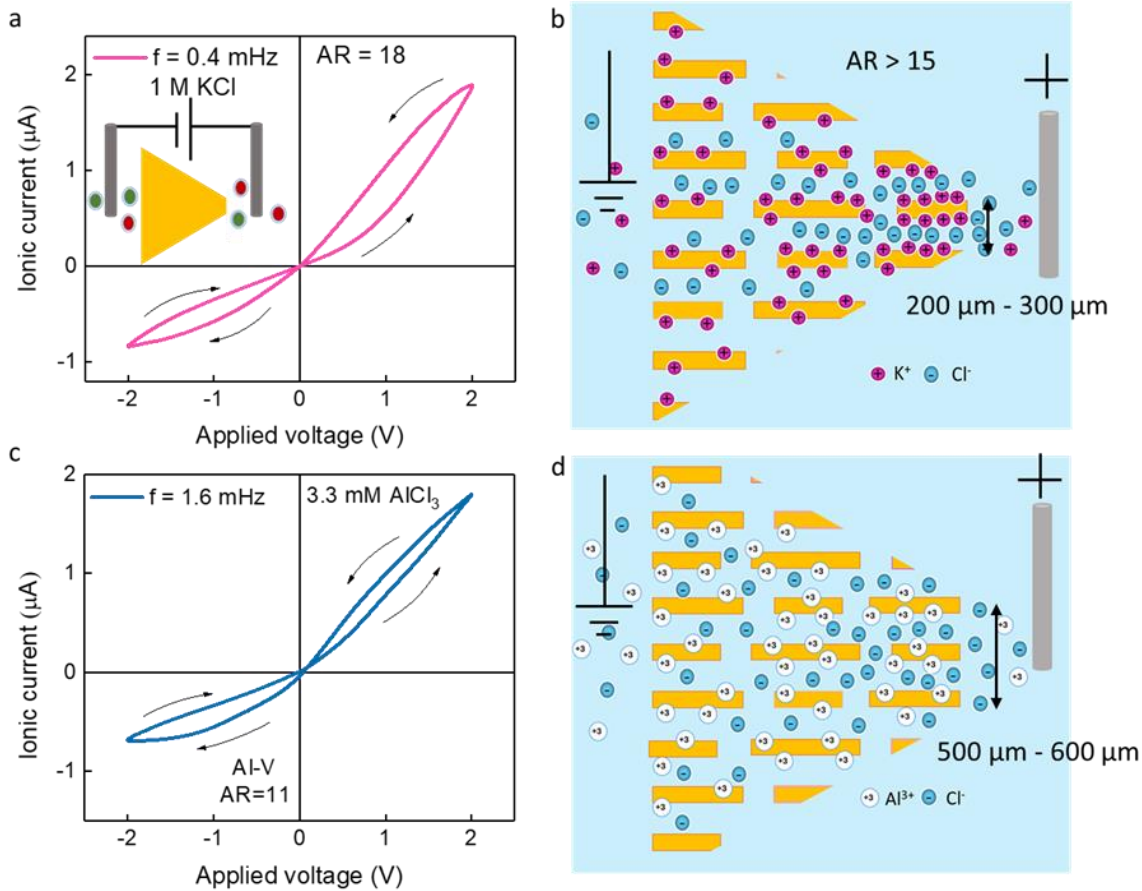

**Figure 3.** Charge inversion effect with monovalent and multivalent ions. (a) The *I-V* curve recorded for a K-V device with an AR of 18 at 0.4 mHz using a 1 M KCl solution shows a pinched and flipped loop. With a positive applied potential, the ionic current increases, indicating that Cl$^-$ is responsible for the ion transport. Inset: Ion transport measurement setup. (b) Schematic of the ion transport mechanism through the K-V samples with AR > 15. It demonstrates that due to the over-screening of K$^+$ ions near the charged surface, the surface charge shifts from negative to positive, allowing Cl$^-$ to pass through the narrow entrance under positive applied potentials. (c) *I-V* curve for an Al-V device with an AR of 11 at 1.6 mHz using a 3.3 mM AlCl$_3$ solution (Note: This data is used in Figure 2c). With a positive applied potential, the current increases, indicating that Cl$^-$ is transported through the channel. (d) Schematic of the ion transport mechanism through the Al-V channels. Multivalent Al$^{3+}$ ions can form a two-dimensional layer near the charged surface, which can switch the surface charge to positive above a certain concentration. As a result, the channels preferentially transport Cl$^-$ ions when a positive potential is applied.

Although SCL theory cannot predict the charge reversal effect observed in our high AR K-V samples, our estimations provide a threshold concentration for KCl as 56 mM. There exist a similar study on mica samples with monovalent rubidium ions, which shows a charge reversal effect above a certain concentration[32]. In the case of mica samples, monovalent ions could trigger inversion due to charge over-screening on the surface. Our K-V samples have transport channels that are extremely confined with a size of 3 Å, and a charge inversion can likely happen. The two-dimensional vermiculite channel is wider on one side and narrower on the other side and the height of the channel is 0.3 nm for K-V samples, which is maintained throughout the channel. This differs from the commonly discussed conical nanopores, where one side has a bigger diameter (base) than the other (tip). In those conical nanopores, obtaining charge inversion was found to be difficult for monovalent ions but plausible with di- and trivalent ions[27, 30]. It is also reported that $SiO_2$ nanochannels with a height of approximately 20 nm able to inverse the surface charge density with divalent ions but failed with monovalent ion[33]. It is possible that in such systems, monovalent ions are unable to form a two-dimensional layer close to the surface due to less confinement of the channel. However, in our K-V laminates, Debye layer overlap occurs even at high ion concentrations of 1 M. Further, our different shaped (symmetric and asymmetric) K-V devices further confirm the flipping of I-V loop (Figure S7a-c). With symmetric devices, I-V shows ohmic characteristics.

The memristive effect with all the salts at different scan frequencies is quantified using the area of the loop enclosed by the I-V curve. From Figure 2a, the area at 0.4 mHz is 0.8 µA V and for 50 mHz it is 0.01 µA V. The area enclosed by the I-V loop decreases with an increase in frequency. As the frequency of the signal increases beyond a threshold, the I-V shows an ohmic behaviour as the ions cannot follow the fast signal. The frequency window for observing the memristor characteristics of our device is from 0.1 to 50 mHz. The pristine $MoS_2$ and activated carbon 2D channel reports an operational frequency ranging from 0.1 to 200 mHz[18].

**Mimicking long-term plasticity patterns:**

We observed that cation-intercalated vermiculite samples exhibit a bipolar memristive effect and retain memory after the applied voltage is removed. Our device can emulate some functionalities of the biological synapses using electrical pulses. The conductance can be increased or decreased in response to a voltage spike, akin to action potentials in biological synapses. For studying long-term potentiation and depression, we selected Ca-V samples with an AR of 12, using 50 mM $CaCl_2$ solution (Figure 4a). To investigate long-term potentiation (an increase in conductance), we applied periodic pulses with a write voltage of +2 V for 56 s and measured the conductance with a read voltage of +10 mV for 56 s. For long-term depression (a decrease in conductance), we applied pulses with a write voltage of -2 V for 56 s and measured the conductance with a read voltage of -10 mV for 56 s. We observed a decrease in current with positive potential and an increase with negative potential, consistent with our I-V characteristics for Ca-V samples at low AR.

We also examined a K-V sample with a 1 M KCl solution and a high AR of 18 (Figure 4b). Here, we observed an increase in current with positive potential and a decrease with negative potential, in agreement with the I-V curves. As discussed previously, $Cl^-$ ions are responsible for ion transport in high AR K-V samples, which is confirmed by our long-term potentiation and depression data. Similar behavior is observed for other cation-intercalated vermiculite membranes.

The relaxation time (t1) is calculated by fitting the Figure 4a&b learning and forgetting process using ∼ exp(-t/t1). From Figure 4a we found that t1 is 2238 s and 1772 s for learning and forgetting process while for Figure 4b, t1 for learning and forgetting process is 1202 s and 1028 s respectively. This indicates that for enrichment of ions (ON state) it takes longer time and for depletion (OFF state) it takes lesser time[34].

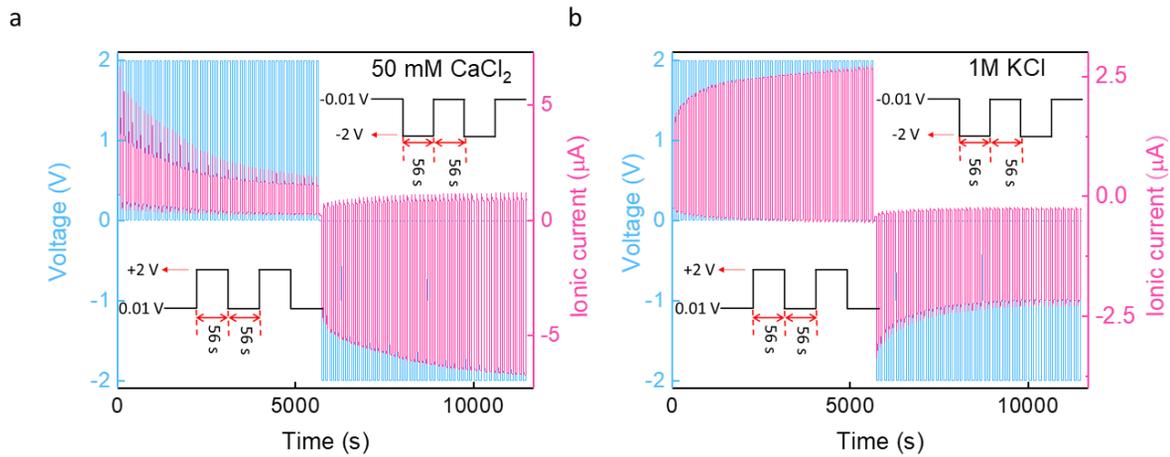

**Figure 4.** Emulating the learning and forgetting process in our angstrom fluidic system. (a) The magnitude of the ionic current change in the Ca-V device as a function of the number of voltage spikes over time, using a 50 mM CaCl₂ solution with an AR of 12. Fifty positive voltage spikes (+2 V, 56 s) were applied, followed by fifty write spikes (-2 V, 56 s) with a small read voltage (±10 mV, 56 s) between each spike. The ionic current decreases with positive voltage spikes and increase with negative spikes. (b) The magnitude of the ionic current change in the K-V device as a function of the number of voltage spikes over time, using a 1 M KCl solution with AR 18. Fifty write spikes (+2 V, 56 s) were applied, followed by fifty erase spikes (-2 V, 56 s) with a small read voltage (±10 mV, 56 s) between each spike. The spike current increases with positive voltage and decreases with negative voltage.

With the help of trapezoidal shaped vermiculite laminates of interlayer spacing 3 to 5 angstroms and with various type of salts, we were able to demonstrate extremely tunable memristor characteristics as per Chua's theory. The memory response time in our in-plane devices is in few minutes, however, it can be further improved by reducing the length of the devices. Instead of in-plane devices, out-of-plane devices can be fabricated, which reduce the transport length. 2D material-based membranes are ideal as it is feasible to make maximally thin membranes. The experiments with different type of membranes helped us to confirm the role of extreme confinement and Debye layer overlap in inducing memristor effect even at 1 M concentration, which is a rare observation. At 1 M concentration, the Debye length is ∼ 0.3 nm, and the Debye layer overlaps for channels of height ≤ 0.6 nm. For channels of height > 0.6 nm, both the cations and the anions could move freely inside the channel. In such cases, an ohmic *I-V* without any memristor behavior was seen. In contrast, our highly confined channels host an overlapped Debye layer, which causes the accumulation of ions at the entrance and/or exit due to channel asymmetry and hence the observation of memristive behaviour at such high concentrations.

Such is the range of laminates that vermiculites offer, which make extreme confinements an interesting avenue for both fundamental and technological advancement. We note that in the case of $AlCl_3$ transport, the bulk mobility of $Cl^-$ is higher than that of $Al^{3+}$, which might argue for an anion-dominated transport, however, the inversion and the memristor effect are crucially dependent on the concentration and also the asymmetry, which argues against the bulk mobility difference as the reason for the flipped *I-V* characteristics.

To emulate the brain's functionality, we studied potentiation and depression of ionic conductance by giving voltage stimulus to the membranes, and each conductance state represents a unique memory state. We tested LTP and LTD with all the salt solutions, however, data for $CaCl_2$ and KCl is presented as the brain utilises solvated $Ca^{2+}$, and $K^+$ ions for most of the tasks. Several other groups also demonstrated memristor characteristics at mHz frequencies arising from slow ion diffusion[18]. To increase the energy efficiency of the devices, we should narrow the voltage range and increase the frequency. Our future work will focus on these aspects.

We also observed that asymmetric geometry is a suitable platform for observing ion-ion interactions and efforts should be made to enhance the AR further. Testing a variety of multivalent ions at salt concentrations beyond 1 M could reveal several ion interaction effects. Laminates made of graphene oxide (GO), MXenes or $MoS_2$ could be explored, additionally. With suitable functional groups, the fluidic channel size can be appropriately tuned in such laminates. Further, with a gate electrode, the type of surface charge can be reversed, providing another parameter for tuning memristor characteristics. By heating these membranes, it is also possible to tune the conductance states.

**Conclusions**

We have demonstrated perfect memristor behavior by creating geometrically asymmetric cation-intercalated vermiculite membranes. The memristor characteristics are highly tunable with external parameters such as frequency, geometry, intercalants, ion concentration and ion type. We observe that charge inversion sometimes occurs, with a change in the nature of the ion transport, resulting in a flip in the memristor behavior. We infer that ion exchange dynamics, ion-ion interactions, and charge accumulation/depletion mechanisms are responsible for the memristor effects. The cost-effective nature of vermiculite membranes makes them attractive for commercial applications compared to expensive 2D material membranes like GO and MXenes. Additionally, vermiculite laminates could be explored for ion correlation effects.

**Methods**

**Materials:** Natural vermiculite crystals (2-3 mm size), potassium chloride, sodium chloride, lithium chloride, calcium chloride, aluminum chloride, and PVDF porous substrate (0.22 μm pore size) were purchased from Sigma Aldrich.

**Fabrication of vermiculite laminates:** We used a two-step ion exchange process described in our earlier study[24], to thermally expand the naturally occurring vermiculite crystals and intercalate them with cations. Essentially, the first step is soaking 100 mg of vermiculite crystals in a 200 mL saturated NaCl solution, refluxing for 24 hours at 100 °C, and then washing the soaked crystals eight to ten times with deionized water to get rid of any remaining salt. $Na^+$ ions are exchanged with interlayer $Mg^{2+}$ cations during this procedure.

The vermiculites that had been sodium exchanged were next immersed in 200 mL of a 2 M lithium chloride (LiCl) solution, refluxed for a further 24 hours, and then again washed with DI water until the excess chloride ions were removed. Following their drying, the resulting Li-exchanged vermiculite (Li-V) crystals were sonicated for 30 min with 1 mg/mL concentration (DI water as solvent). Either the solution centrifuged for 15 minutes at 3000 rpm or letting the thicker flakes settle naturally under gravity overnight gives us a dispersion of monolayers flakes. The vermiculite membrane was finally prepared using a vacuum filtration assembly with the supernatant, which comprises monolayer. PVDF with a pore size of 0.22 μm was employed as a supporting membrane for the creation of a vermiculite membrane, which was readily peeled off after drying the prepared membrane under IR lamp for 15 mins. A 24-hour immersion in 1 M KCl, NaCl, CaCl$_2$, and AlCl$_3$, makes these free-standing membranes intercalated with various cations, including $K^+$, $Na^+$, $Ca^{2+}$, and $Al^{3+}$. According to our earlier study[24], vermiculite membranes are extremely water stable because of the exchanged cations. After rinsing it with water, the cation-exchanged membrane was dried. These membranes were used for the device fabrication and further ion transport studies. '

**Supporting information**

XRD analysis of cation intercalated membranes (Figure S1), Digital photo of membrane (Figure S2a), SEM cross-sectional image (Figure S2b), Zeta potential measurement (Figure S3), flipping of *I-V* curve with concentration (Figure S4 a-d), flipped *I-V* curve at high AR for KCl and NaCl (Figure S5a&b), charge inversion effect with low and high AR for NaCl (Figure S6a&b), shape controlled ICR for KCl (Figure S7a-c), surface charge density calculation of the nanosheets.

**Authors contribution**

D.B., and K.G. conceived the idea. D.B. designed the project, fabricated the devices, performed ion transport measurement, sample characterization, and data analysis. P.A. fabricated the devices, performed ion transport measurement and data analysis. L.H.Y. helped in data analysis. D.B., P.A., and K.G. wrote the manuscript. K.G. supervised the project and acquired the fund. All the authors discussed the results and commented on the manuscript.

**Acknowledgments**
This work was mainly funded by DST-INAE with grant no. 2023/IN-TW/09 and supported by MHRD STARS with grant no. MoE-STARS/STARS-1/405 and Science and Engineering Research Board (SERB), Government of India, through grant no. CRG/2019/002702. K.G. acknowledges the support of Kanchan and Harilal Doshi chair fund. P.A. acknowledges the Sabarmati Bridge Fellowship of IIT Gandhinagar. The authors acknowledge the contribution from IITGN central instrumentation facility. L.H.Y. acknowledges the financial supports from the National Science and Technology Council (NSTC), Taiwan under Grant No. NSTC 112-2923-E-011-003-MY3, 113-2628-E-011-002, 113-2628-E-011-005-MY3, and 113-2124-M-011-002, and from the Ministry of Education of Taiwan (MOE, "Sustainable Electrochemical Energy Development Center" (SEED) project).

**X-ray diffraction (XRD) analysis:**

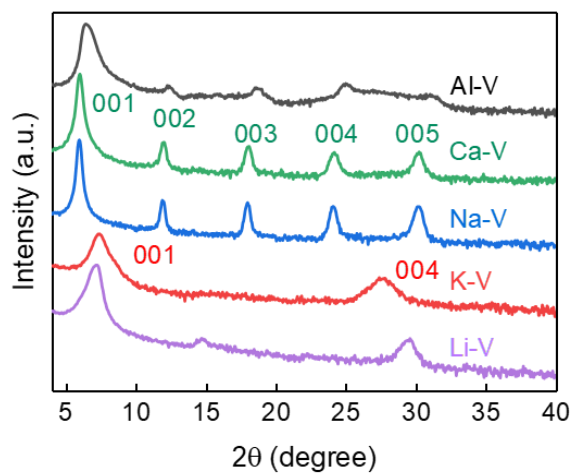

**Figure S1.** XRD data of cation-intercalated vermiculite membrane. The measurement of the cation-intercalated vermiculite was carried out using a Multipurpose XRD instrument, Rigaku, Japan. The XRD data shows that as the cations are intercalated, there is a shift in the interlayer spacing related to the intercalation of one or two layers of water molecules. After removing a unit thickness of vermiculite crystal i.e., 9.6 Å, we found that the free interlayer spacings left for K-V, Na-V, Ca-V and Al-V samples are ~3 Å, ~5 Å, ~5 Å, and ~5 Å respectively.

**Microstructure and Thickness of Membranes:**

a

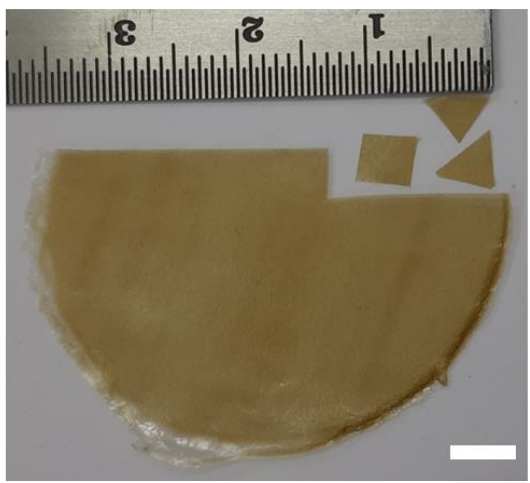

b

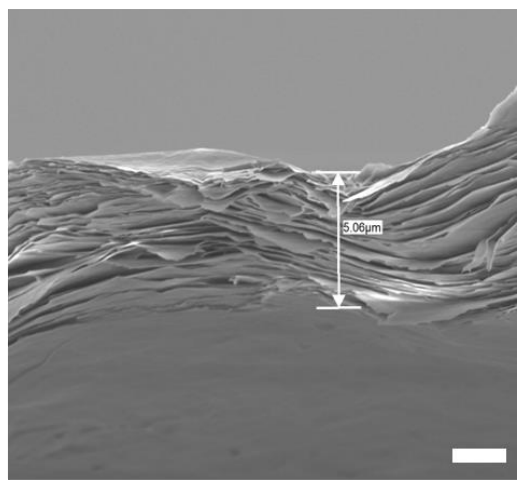

**Figure S2.** SEM characterization. (a) Free-standing vermiculite membrane. The scale bar is 5 mm. (b) SEM cross-sectional image of the vermiculite membrane shows laminated structure, and the measured thickness is ~5 μm. The scale bar is 2 μm.

**Zeta potential measurement of intercalated vermiculite dispersions:**

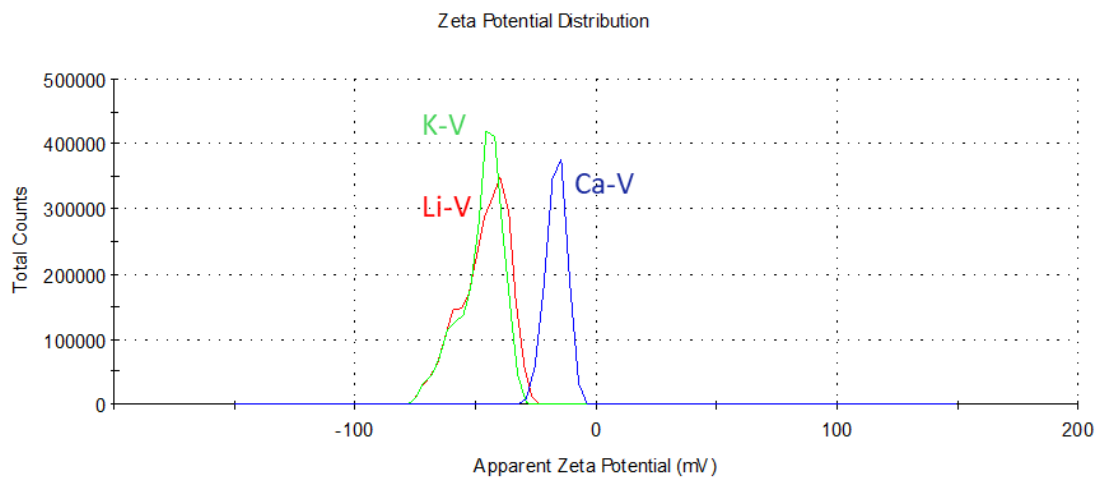

**Figure S3.** Zeta potential of the cation-intercalated vermiculite dispersions. The dispersed solution of vermiculite with different intercalants was prepared with a concentration of 1 mg/ml for the zeta potential measurement (Malvern Zetasizer ZEN3600, UK). The negative value of the zeta potential of Li-V, K-V, and Ca-V membranes suggests a negative surface charge. As the valence of intercalant cation increases, the zeta potential decreases.

**Evolution of flipped *I-V* characteristics with concentration:**

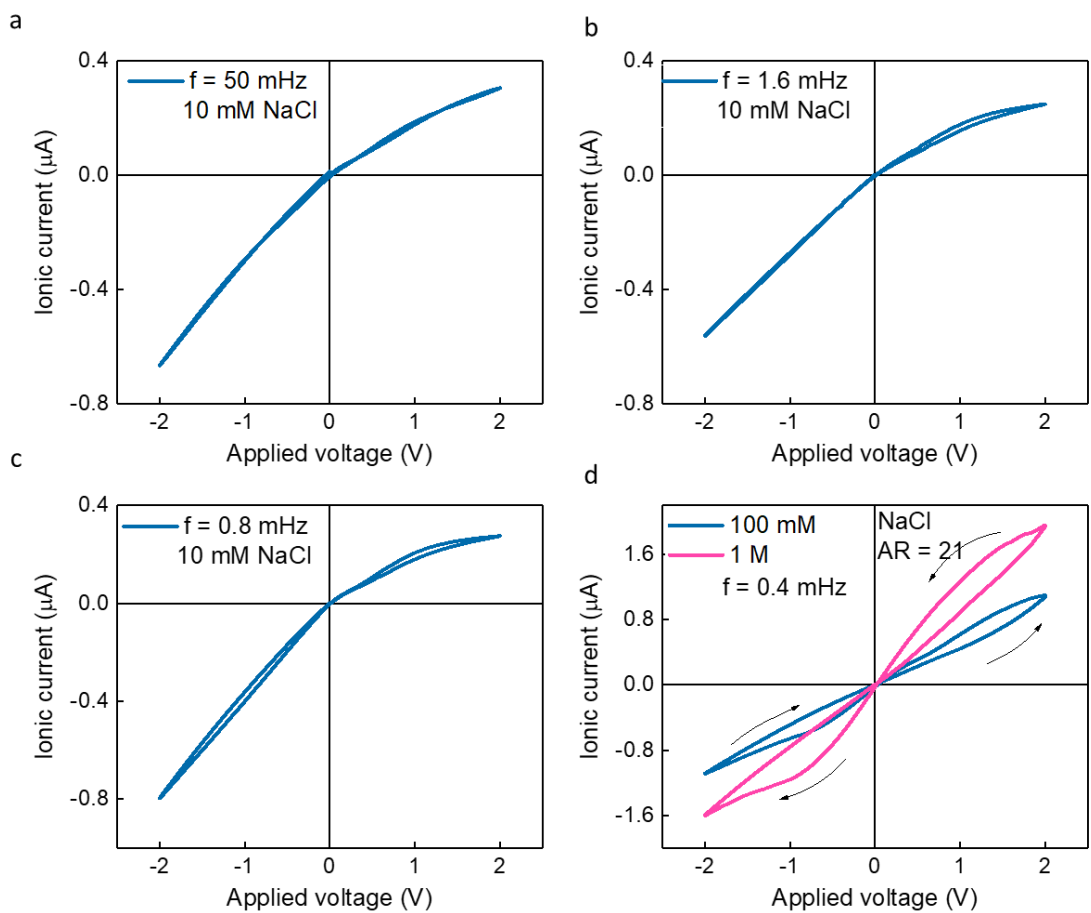

**Figure S4.** Flipped *I-V* characteristics of Na-V membrane of AR 21. (a) *I-V* curve shows typical ion current rectification (ICR) with 10 mM NaCl at 50 mHz. (b) *I-V* curve for 10 mM NaCl at 1.6 mHz. (c) Pinched *I-V* loop with 10 mM NaCl at 0.8 mHz. (d) ICR equals to 1 with 100 mM NaCl at 0.4 mHz and ICR of 0.7 with 1 M NaCl at 0.4 mHz. The *I-V* curve shows the evolution from an ICR to an inverted ICR with an increase in concentration at different frequencies. The magnitude of ICR is estimated from the ratio of ionic current at -2 V to that at +2 V.

**Flipped *I-V* curve with 1 M KCl and 1 M NaCl for devices of AR above 15 at different frequencies:**

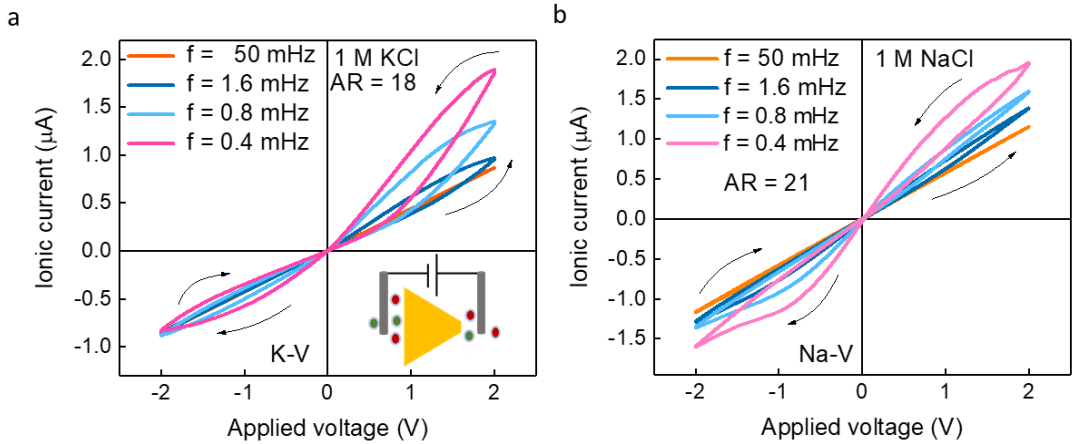

**Figure S5.** Flipped *I-V* curve at 1 M concentration for K-V and Na-V device. (a) *I-V* curve of K-V device with 1 M KCl at different frequencies. For this K-V device, we choose an AR of 18. Compared to devices with AR < 15, these devices show flipped *I-V* behavior. Bottom Inset: schematic of ion transport measurement setup. (b) *I-V* curve of Na-V devices with 1 M NaCl at different frequencies. For this Na-V device, we choose an AR of 21. The *I-V* curve shows memristor characteristics at 1 M concentration for both KCl and NaCl solutions. With the decrease in frequency, the conductance increases. (Note: *I-V* curve of K-V device at *f* = 0.4 mHz with 1 M KCl solution and an AR of 18 has been provided in the main text as Figure 3a).

**Charge inversion with asymmetric ratio with Na-V device and NaCl solutions:**

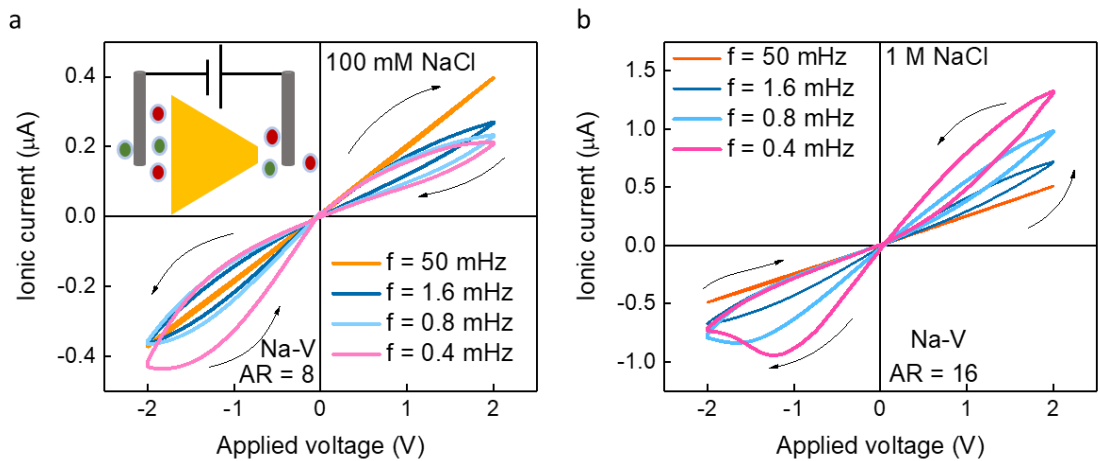

**Figure S6.** Memristive effect of Na-V samples with NaCl. (a) *I-V* curves at different scanning frequencies at 100 mM NaCl with an AR of 8. (b) *I-V* curve at different scanning frequencies at 1 M NaCl with an AR of 16. With two different asymmetric ratios (AR) of 8 and 16, the *I-V* characteristic looks similar to that of K-V samples. The *I-V* gets flipped when the sample asymmetric ratio (AR) is increased beyond 15.

**Ion transport through symmetric and asymmetric K-V devices:**

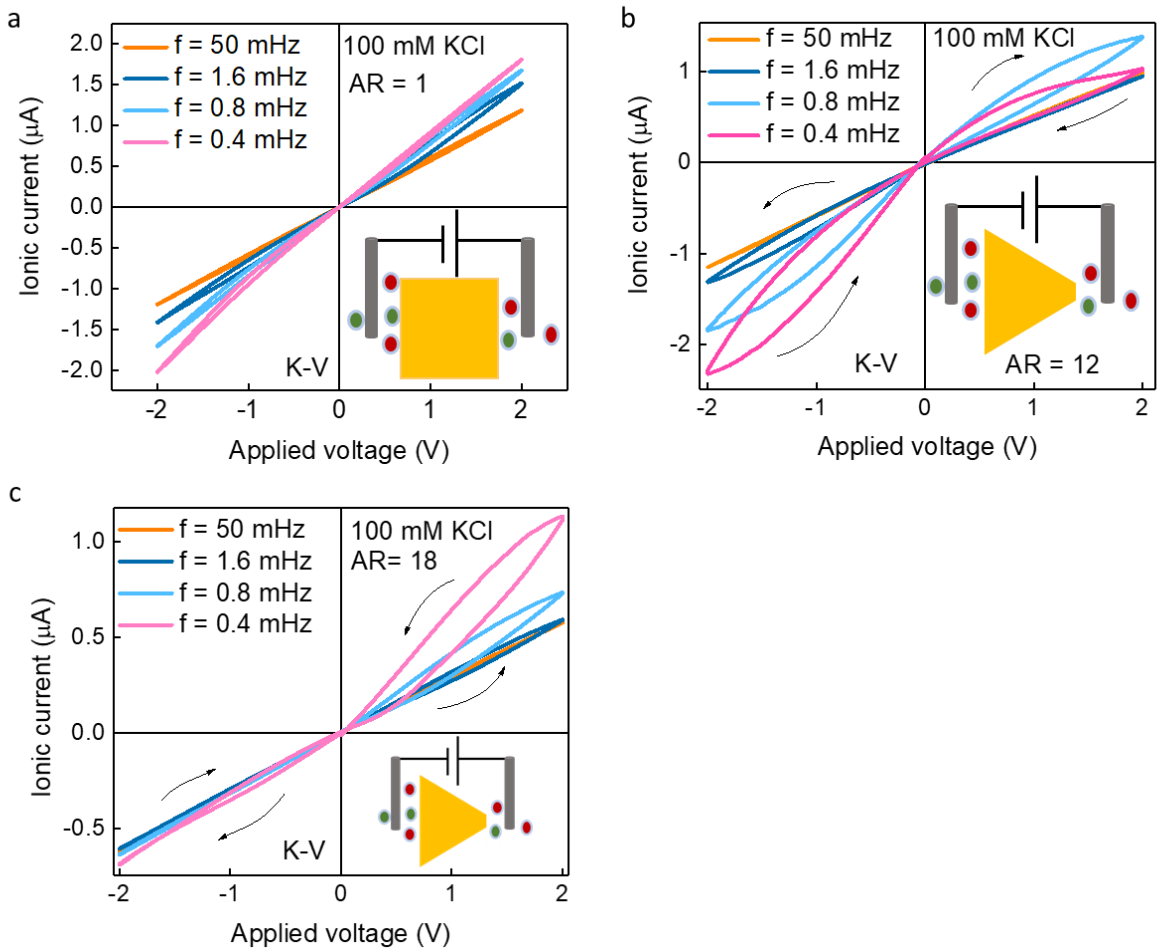

**Figure S7.** Ion transport across symmetric and asymmetric K-V devices. (a) I-V curves at different scanning frequencies at 100 mM KCl for K-V samples with AR of 1. This shows ohmic *I-V* with ICR close to 1. (b) *I-V* curves at different scanning frequencies at 100 mM for K-V samples with AR of 12. This shows an ICR ($I_{-2V}/I_{+2V}$) of 2.25 at 0.4 mHz. (Note: This data is used in the main text in Figure 2a). (c) *I-V* curves at different scanning frequencies at 100 mM for K-V samples with AR of 18. This shows an ICR ($I_{-2V}/I_{+2V}$) of 0.6 at 0.4 mHz.

**Supplementary Note 1 - Surface charge density calculation:**

Vermiculite possesses a negative surface charge due to the substitution of $Al^{3+}$ impurities at $Si^{4+}$ octahedral sites. These excess charges are balanced by the exchangeable cations. From the zeta potential measurement, we found that the K-V has a potential of -48 mV. According to Gouy-Chapman equation[1], the surface charge density, $\sigma$, is estimated from the measured zeta potential, $\xi$, which is shown below (eq S1);

$$\sigma = -\frac{\varepsilon_0 \; \varepsilon_r \xi}{\lambda_d} \left( \frac{\sinh\left(\frac{F\xi}{2RT}\right)}{\frac{F\xi}{2RT}} \right) \tag{S1}$$

where, $\varepsilon_r$ is the dielectric constant, $\varepsilon_0$ is the permittivity of the free space, $\lambda_d$ is the Debye length, $F$ is the Faraday constant, $R$ is the gas constant, and $T$ is the temperature. The surface charge is found to be $\sim$-4 mC/m², assuming $\varepsilon_r$ = 80.